\documentclass[preprint, trackchanges]{aastex61}
\usepackage{graphicx, graphics, amsmath,amssymb, amsfonts}%Include figure fiels
\newcommand{\gsim}{\mbox{\hspace{.2em}\raisebox{.5ex}{$>$}\hspace{-.8em}\raisebox{-.5ex}{$\sim$}\hspace{.2em}}}
\newcommand{\lsim}{\mbox{\hspace{.2em}\raisebox{.5ex}{$<$}\hspace{-.8em}\raisebox{-.5ex}{$\sim$}\hspace{.2em}}}

\newcommand{\E}[1]{\times 10^{#1}}
\newcommand{\twCO}{$^{12}$CO}  \newcommand{\thCO}{$^{13}$CO}

\newcommand{\HII}{\mbox{H\,\textsc{ii}}}

      \newcommand{\ps}{\,{\rm s}^{-1}}
       
    \newcommand{\km}{\,{\rm km}}

\begin{document}

\title{
Is HESS J1912$+$101 Associated with an Old Supernova Remnant?
}

\shorttitle{Is HESS J1912$+$101 associated with an old SNR}

\correspondingauthor{Yang Su}
\email{yangsu@pmo.ac.cn}

\author[0000-0002-0197-470X]{Yang Su}
\affil{Purple Mountain Observatory and Key Laboratory of Radio Astronomy,
Chinese Academy of Sciences, Nanjing 210008, China}

\author{Xin Zhou}
\affiliation{Purple Mountain Observatory and Key Laboratory of Radio Astronomy,
Chinese Academy of Sciences, Nanjing 210008, China}

\author{Ji Yang}
\affiliation{Purple Mountain Observatory and Key Laboratory of Radio Astronomy,
Chinese Academy of Sciences, Nanjing 210008, China}

\author{Yang Chen}
\affiliation{Department of Astronomy, Nanjing University,
Nanjing 210023, China}
\affiliation{Key Laboratory of Modern Astronomy and Astrophysics,
Nanjing University, Ministry of Education, Nanjing 210023, China}

\author{Xuepeng Chen}
\affiliation{Purple Mountain Observatory and Key Laboratory of Radio Astronomy,
Chinese Academy of Sciences, Nanjing 210008, China}

\author{Yan Gong}
\affiliation{Purple Mountain Observatory and Key Laboratory of Radio Astronomy,
Chinese Academy of Sciences, Nanjing 210008, China}

\author{Shaobo Zhang}
\affiliation{Purple Mountain Observatory and Key Laboratory of Radio Astronomy,
Chinese Academy of Sciences, Nanjing 210008, China}

\begin{abstract}
HESS J1912$+$101 is a shell-like TeV source that has no clear counterpart 
in multiwavelength. Using CO and \mbox{H\,\textsc{i}} data, we reveal that 
$V_{\rm LSR}\sim+60$~$\km\ps$ molecular clouds (MCs), together with shocked 
molecular gas and high-velocity neutral atomic shells, are concentrated 
toward HESS J1912$+$101.
The prominent wing profiles up to $V_{\rm LSR}\sim +80\km\ps$ seen in 
\twCO\ ($J$=1--0 and $J$=3--2) data, as well as the high-velocity expanding 
\mbox{H\,\textsc{i}} shells up to $V_{\rm LSR}\sim +100\km\ps$, exhibit
striking redshifted-broadening relative to the quiescent gas. 
These features provide
compelling evidences for large-scale perturbation in the region.
We argue that the shocked MCs and the high-velocity 
\mbox{H\,\textsc{i}} shells may originate from an old 
supernova remnant (SNR). The distance to the SNR is
estimated to be $\sim4.1$~kpc based on the \mbox{H\,\textsc{i}} 
self-absorption method, which leads to a physical radius of 29.0~pc 
for the $\sim$(0.7--2.0)$\times10^{5}$~years old remnant with an expansion 
velocity of $\gsim40$~km~s$^{-1}$. 
The +60~km~s$^{-1}$ MCs and the disturbed gas are indeed found to 
coincide with the bright TeV emission, 
supporting the physical association between them. 
Naturally, the shell-like TeV emission comes from the decay of 
neutral pions produced by interactions between the accelerated hadrons from
the SNR and the surrounding high-density molecular gas.
\end{abstract}

\keywords{ISM: individual objects (shell-like clouds, HESS J1912$+$101)
-- ISM: kinematics and dynamics -- ISM: supernova remnants}

\section{INTRODUCTION}
Shell-type supernova remnants (SNRs) are believed to be efficient
accelerators of high-energy cosmic rays (CRs) through 
diffusive shock acceleration 
\citep[e.g., see the recent review of][]{2008ARAA4689R}.
Such strong shocks of young SNRs put considerable energy into 
accelerated particles of CRs, resulting in 
high-energy X-ray and $\gamma$-ray emission.
On the other hand, pp-interactions and the subsequent decay of 
neutral pions will significantly enhance the $\gamma$-ray emission
in high-density environments. 
Molecular clouds (MCs) near SNRs are thus promising sources of 
$\gamma$-ray emission because of their high densities. 
Recently, theoretical 
\citep[e.g.,][]{2006MNRAS.371.1975Y,2009ApJ...707L.179F,2010MNRAS.409L..35L,
2011MNRAS.410.1577O} and observational studies 
\citep[e.g.,][]{2008AA...481..401A,2008ApJ...679L..85T,2010ApJ...717..372C,
2010ApJ...710L.151T,2013Sci...339..807A} 
have been performed to investigate the high-energy $\gamma$-ray
emission toward SNRs and their ambient molecular gas
\citep[e.g., see a summary in][]{2015SSRv..188..187S}.

HESS J1912$+$101 is an extended TeV source that was discovered
by \cite{2008AA...484..435A} and displays the prominent shell 
appearance of the TeV emission from the further increased H.E.S.S. 
exposure \citep{2015ICRC...34..886P}. 
However, no firm conclusion on the nature of the source can be drawn,
due to the lack of the counterpart of HESS J1912$+$101 at other wavebands.
The TeV source was speculated
to associate with a high spin-down luminosity pulsar PSR J1913$+$1011
or an old SNR \citep{2008AA...484..435A}. 
The pulsar wind nebulae (PWN) scenario may be excluded because of 
no detectable X-ray emission around PSR J1913$+$1011
and/or the geometrical centroid of HESS J1912$+$101
\citep{2008ApJ...682.1177C} and the presence of the TeV shell-like
structure.
Therefore, the shell-like source of HESS J1912$+$101 is probably a TeV 
SNR candidate \citep{2015ICRC...34..886P,2016arXiv161200261G}, 
although no corresponding radio counterpart has been found to 
match the TeV shell structure yet. 

The high angular resolution of the very-high-energy 
(VHE, $E>$100~GeV) data, which is better than
0\fdg1 for the H.E.S.S. observation, allows us to compare the TeV
structure with that seen in other wavelengths with the precision of several
arcmin. Because little corresponding radio and X-ray emission matches
the high-energy source, typical structures of the molecular and/or atomic 
gas on a large scale (e.g., shells, arcs, and bubbles, etc.), together with 
possible kinematic signatures of the disturbed gas (e.g., molecular line
broadenings, high-velocity structures, and large-velocity gradients, etc.), are
useful and important in investigating the association 
between HESS J1912$+$101 and its environment.

In this paper, we present the molecular and atomic line study toward 
the shell-like TeV source of HESS J1912$+$101. 
Both of CO and \mbox{H\,\textsc{i}} datasets reveal the partial 
shell-like structures and high-velocity gas emission in the field of 
view (FOV) of the TeV source, indicating an energetic source with high 
momentum and mechanical energy injection there. We suggest that the 
energetic source is an old SNR that is interacting with the 
surrounding $V_{\rm LSR}\sim+60$~km~s$^{-1}$ giant MC (GMC)
located at a near kinematic distance of 4.1 kpc.
Accordingly, the shell-like TeV $\gamma$-ray emission
of HESS J1912$+$101 arises from the interaction between the 
(0.7--2.0)$\times10^{5}$~years old SNR and its ambient dense molecular gas.

This paper is structured as follows.
In Section 2, we briefly describe the CO observation, the data 
reduction, and other survey data used in the paper. 
In Section 3, we give our results of the shocked molecular gas and
the high-velocity \mbox{H\,\textsc{i}} shells; investigate
the association between the disturbed gas and an old SNR; and
discuss the origin of the shell-like TeV emission of HESS J1912$+$101. 
Finally, Section 4 summarizes the results.

\section{CO and \mbox{H\,\textsc{i}} data}
CO datasets are from the Milky Way Imaging Scroll Painting 
(MWISP\footnote{http://english.dlh.pmo.cas.cn/ic/}, i.e., \twCO~($J$=1--0), 
\thCO~($J$=1--0), and C$^{18}$O~($J$=1--0) lines) and CO High-Resolution 
Survey \citep[COHRS, \twCO~($J$=3--2) line;][]{2013ApJS..209....8D}.
The \mbox{H\,\textsc{i}} and 21~cm continuum emission data 
are from the VLA Galactic Plane Survey \citep[VGPS;][]{2006AJ....132.1158S}.

The MWISP project is a large, unbiased, and highly sensitive CO, and 
its isotopes survey toward the Galactic plane using the 13.7~m 
millimeter-wavelength telescope located
at Delingha, China. A Superconducting Spectroscopic Array Receiver 
(SSAR) system with a $3\times3$ beam array \citep{Shan} was used as the front end 
and a Fast Fourier Transform Spectrometer (FFTS) with a total bandwidth of 
1 GHz was used as the back end. The half-power beamwidth (HPBW)
of the telescope is about $50''$ for the three lines. 
The typical rms noise level is about 0.5 K for \twCO\ ($J$=1--0) at the
channel width of 0.16$\km\ps$ and 0.3~K for \thCO\ ($J$=1--0)
and C$^{18}$O ($J$=1--0) at 0.17$\km\ps$. 
The final cube data were constructed with a grid spacing of $30''$.
All data were reduced using the 
GILDAS software\footnote{http://ascl.net/1305.010 or
http://www.iram.fr/IRAMFR/GILDAS}.

The information of the 13.7 m telescope can be seen from the status 
report\footnote{http://www.radioast.nsdc.cn/mwisp.php}.
The detailed observing strategy, the instrument, and the quality
of the CO observations can be found in our 
paper published recently
\citep[see Section 2 in][]{2017ApJ...836..211S}. 
We summarize the data used in Table~1.

\section{RESULTS AND DISCUSSIONS}
\subsection{CO emission and the shocked molecular gas}
HESS J1912$+$101 is a shell-like TeV source that
is centered at ($l=$~44\fdg46, $b=-$0\fdg13) with an outer
radius of 0\fdg49 and an inner radius of 
0\fdg32 \citep{2016arXiv161200261G}.
Neither radio nor X-ray emission was detected toward
the shell-like TeV source. 
Due to the significant extinction in the direction of the
inner Galactic plane, optical observations are difficult
to establish the nature of the source.
Therefore, the submillimeter, millimeter, and
centimeter emission is probably helpful in studying
the nature of the counterpart of HESS J1912$+$101.

Using the high-quality CO data, we find that molecular gas 
at a systemic velocity of $V_{\rm LSR}\sim$+60$\km\ps$ is 
concentrated toward the extended emission of the TeV source.
Generally, \twCO\ emission, which is widely distributed in the FOV,
is more diffuse and extended than that of \thCO.
After carefully checking the CO data channel by channel,
we distinguish several molecular partial shell-like structures
toward the TeV source of HESS J1912$+$101. These partial 
shells are relatively prominent in \thCO\ emission,
which exhibits distinct velocity-coherent structure over the long
curved line (length$\gsim 10'$ and width$\lsim 2'$; 
see the thick-dashed gold lines in Figure \ref{f2}).
Meanwhile, optically thin C$^{18}$O emission is also detected 
in some regions of partial shells, indicating the high-density 
there due to its rarer abundance.

\begin{figure}[ht]
\plotone{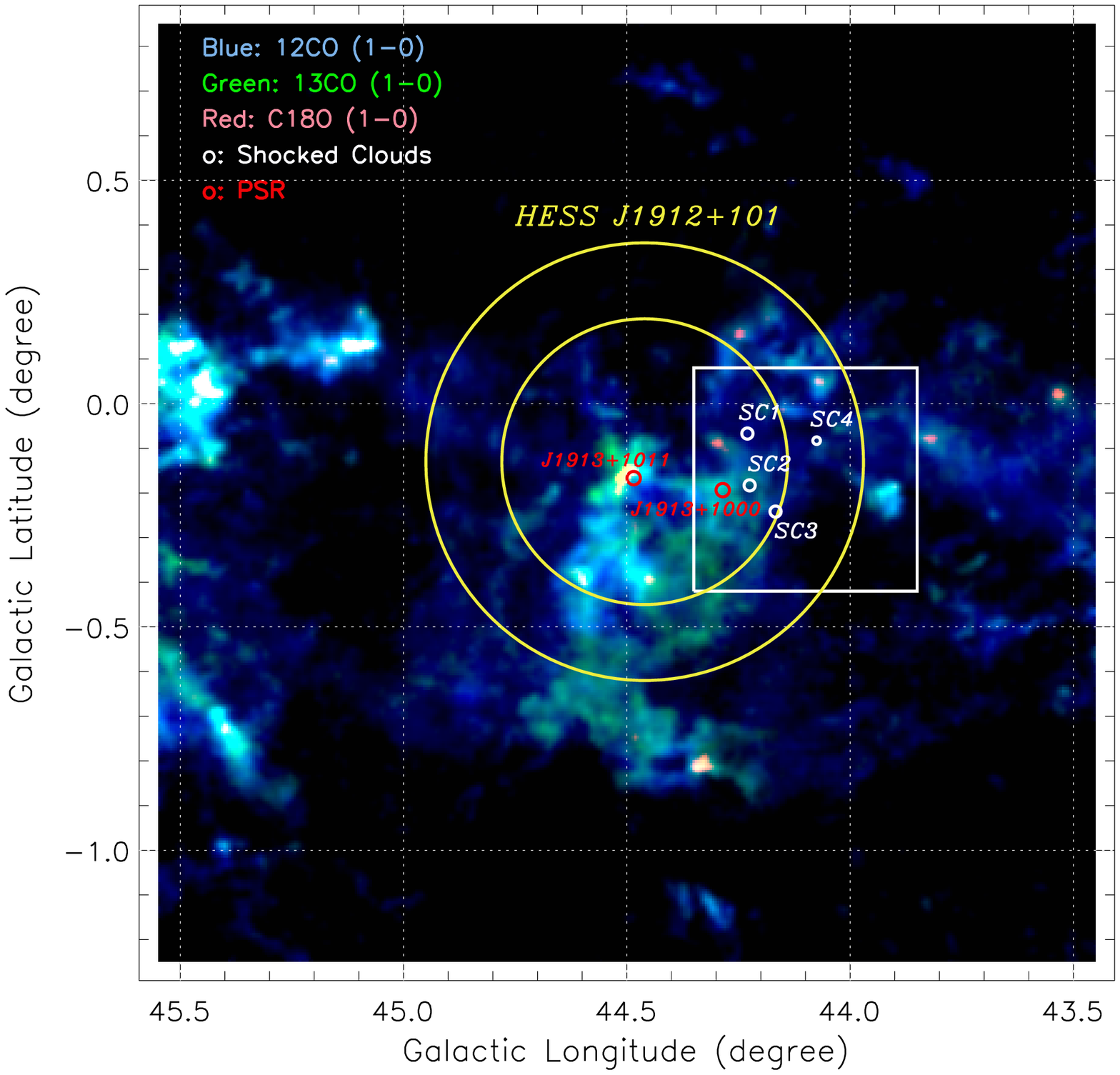}
\caption{\twCO\ ($J$=1--0, blue), \thCO\ ($J$=1--0, green), and
C$^{18}$O ($J$=1--0, red) intensity map
toward HESS J1912$+$101 in the 58.5--62.0~km~s$^{-1}$ interval.
The two yellow circles indicate the outer and inner TeV 
shells of HESS J1912$+$101
\citep{2008AA...484..435A,2016arXiv161200261G}, while the red and the
white circles indicate the two pulsars \citep{2002MNRAS.335..275M,2004MNRAS.352.1439H}
and the shocked clouds. The white box indicates the region shown in Figure 2.
\label{f1}}
\end{figure}

\begin{figure}
\plotone{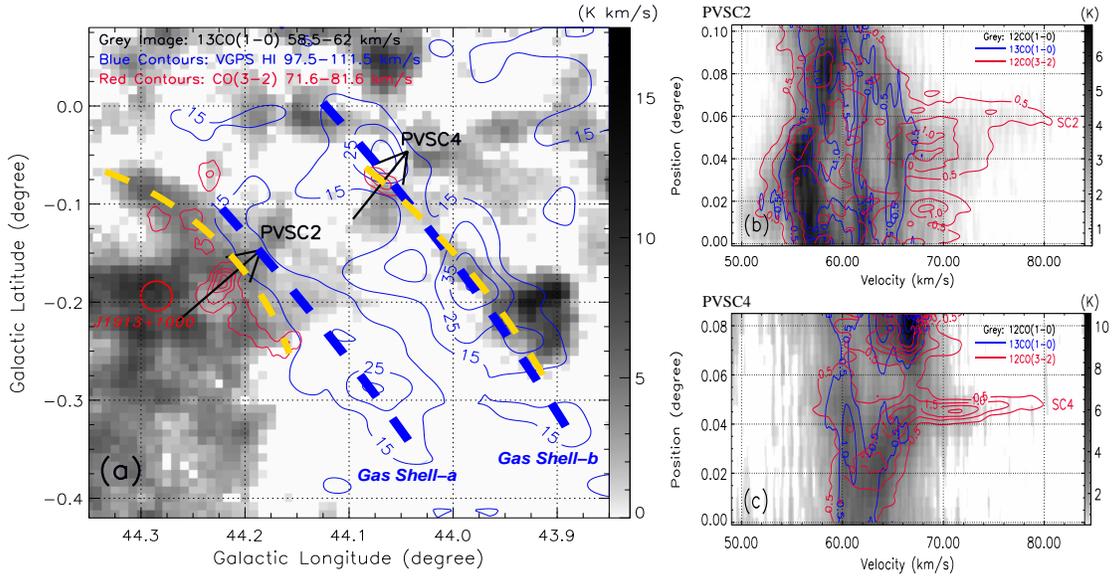}
\caption{ (a) \thCO\ ($J$=1--0) emission in the interval of 58.5--62.0~km~s$^{-1}$,
overlaid with the high-velocity \mbox{H\,\textsc{i}} gas (blue, 
97.5--111.5~km~s$^{-1}$) and the shocked CO ($J$=3--2) gas (red, 
71.6--81.6~km~s$^{-1}$). The thick-dashed blue and gold lines represent 
the shells traced by the high-velocity \mbox{H\,\textsc{i}} gas and 
the unperturbed \thCO\ gas, respectively. The two arrows, which just 
pass through 
the shocked BML regions of SC2 and SC4, 
indicate PV slices shown in panels (b) and (c), respectively.
(b) PV diagram of \twCO\ ($J$=1--0) emission along the arrow of PVSC2
overlaid with the blue contours of \thCO\ emission and the red contours
of CO ($J$=3--2) emission. The PV slice has a length of 6\farcm4
(from ($l=$44\fdg264, $b=-$0\fdg219) to ($l=$44\fdg185, $b=-$0\fdg147))
and a width of 1\farcm5.
(c) PV diagram of \twCO\ ($J$=1--0) emission along the arrow of PVSC4.
The PV slice has a length of 5\farcm2
(from ($l=$44\fdg096, $b=-$0\fdg115) to ($l=$44\fdg044, $b=-$0\fdg046))
and a width of 1\farcm5.
\label{f2}}
\end{figure}

Figure \ref{f1} shows the three-color intensity image of CO emission 
(\twCO\ in blue, \thCO\ in green, and C$^{18}$O in red; 
$V_{\rm LSR}$=58.5--62.0~$\km\ps$) toward the TeV source. 
Assuming the \thCO\ abundance of
$N$(H$_2$)/$N(^{13}$CO) $\approx7\E{5}$ \citep{1982ApJ...262..590F}
and the C$^{18}$O abundance of
$N$(H$_2$)/$N$(C$^{18}$O) $\approx7\E{6}$ \citep{1995A&A...294..835C},
the H$_2$ number density of the dense part of the molecular shells 
traced by the optically thin \thCO\ emission (denoted by the
thick-dashed gold lines in Figure \ref{f2}) is $\sim$400--1600~cm$^{-3}$ at a distance 
of 4.1~kpc (see Section 3.3) with the shell's width of $\sim2'$. 
Adopting the mean CO-to-H$_2$ mass
conversion factor $X_{\rm CO}$=$2\E{20}$~cm$^{-2}$K$^{-1}$km$^{-1}$s
\citep{2013ARAA..51..207B}, the total mass of the molecular partial 
shells traced by the optically thick \twCO\ emission is about several 
thousands solar mass at a distance of 4.1 kpc.
The total mass estimated from \thCO\ and C$^{18}$O emission is somewhat
less than that from \twCO\ emission.
The mean volume density of the $V_{\rm LSR}$=+60~$\km\ps$ MCs is 
approximately several hundreds per cm$^{3}$, while the density in some 
dense regions should be $> 10^3$~cm$^{-3}$.

\begin{figure}
\plotone{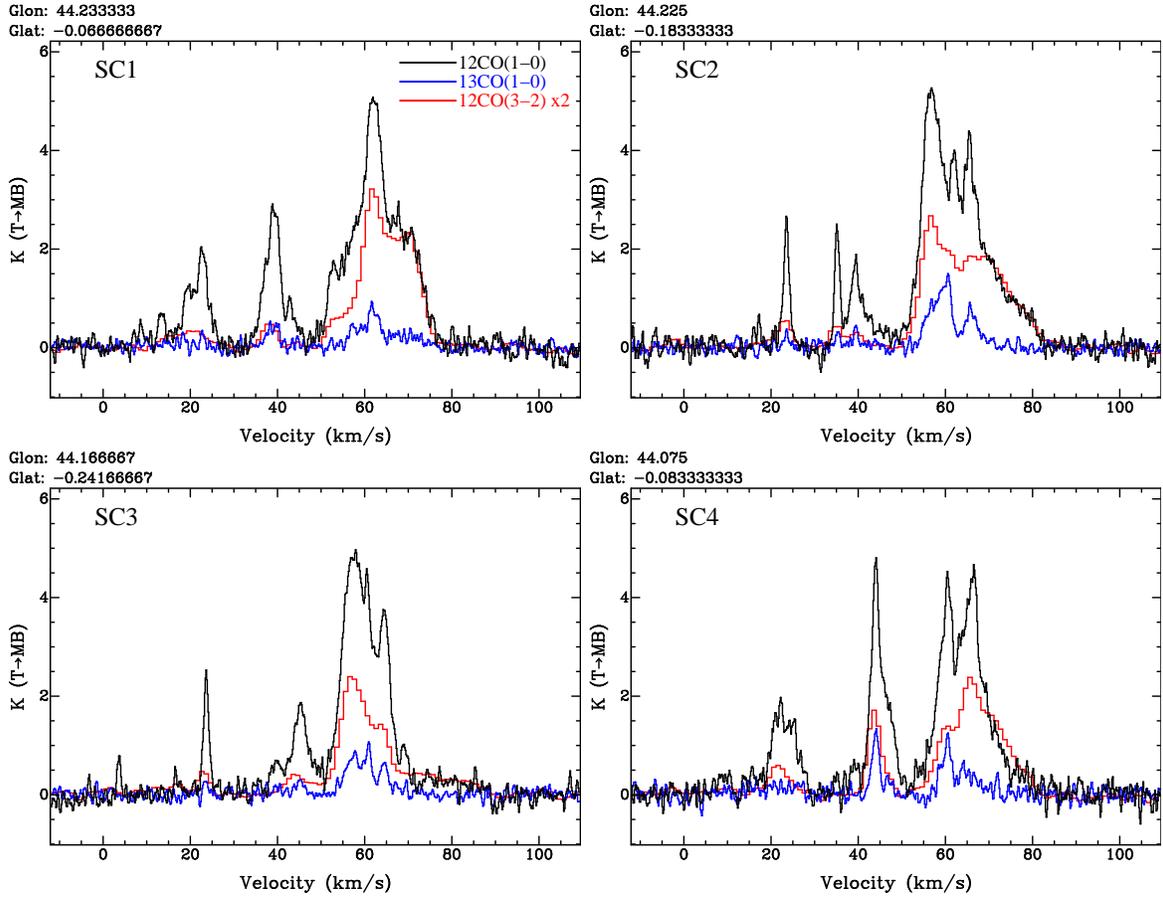}
\caption{\twCO\ ($J$=1--0; black), \thCO\ ($J$=1--0; blue), and
convolved \twCO\ ($J$=3--2; red) spectra of
shocked gas toward HESS J1912$+$101. Positions of shocked BML regions
are labeled in Figure \ref{f1}.
These spectra are extracted from regions of 1\farcm5$\times$1\farcm5,
1\farcm5$\times$1\farcm5, 1\farcm5$\times$1\farcm5, and
1\farcm0$\times$1\farcm0 around points SC1--SC4, respectively.
\label{f3}}
\end{figure}

Furthermore, in combination with the MWISP and COHRS data, we identify 
several shocked regions projected toward the interior of the TeV
shell of HESS J1912$+$101 (see the four white circles of SC1--SC4 in Figure \ref{f1}). 
By comparing the broadened \twCO\ ($J$=1--0 and 3--2) lines
with the narrow \thCO\ line, we can illustrate that the molecular gas
in some regions exhibits striking redshifted-broadening with respect to
the ambient MCs at $V_{\rm LSR} \sim$+60~$\km\ps$.
As shown in Figures \ref{f2}, the broadened
red wing of \twCO\ ($J$=3--2) emission is obviously
up to $V_{\rm LSR}\sim +80\km\ps$ from position$-$velocity (PV) diagrams
of SC2 and SC4. The typical spectra of 
the four regions, which display prominent 
redshifted-broadenings in \twCO\ ($J$=1--0 and 3--2) emission,
are shown in Figure \ref{f3}.

Broadened profiles of \twCO\ emission usually come from the turbulent molecular gas
that is readily and significantly influenced by local shocks.
On the contrary, the 
optically thin \thCO\ emission is mainly from the quiescent and 
unperturbed MCs. Therefore, such features like broadenings, wings, 
or asymmetries in \twCO\ spectra with respect to \thCO\ line show 
convincing evidence for shocked and disrupted molecular material,
which often occurs in SNR--MC interacting systems
\citep[e.g.,][]{1998ApJ...508..690F,2005ApJ...618..297R,
2010ApJ...712.1147J,2016ApJ...816....1K}.

SC2, which is likely a part of the shocked broad molecular line (BML) 
shell (see the red contours near the arrow of PVSC2 in Figure \ref{f2}),
exhibits the association between the shocked medium at
$V_{\rm LSR}$=65--80~$\km\ps$ from the \twCO\ ($J$=1--0 and 3--2) emission 
and the surrounding quiescent MCs at $V_{\rm LSR}$=+59.6~$\km\ps$ 
from the \thCO\ emission (see Figure \ref{f2}).
A piece of shocked gas or SC4 
identified from broadened \twCO\ ($J$=3--2) emission 
(2\farcm0$\times$0\farcm5, see the red contours near the arrow of PVSC4 in Figure \ref{f2})
also displays the
spatial coincidence with the nearby \thCO\ partial shell structure
at the peak velocity of $V_{\rm LSR}$=+60.3~$\km\ps$. 
We make a PV diagram perpendicular to
the main axis of SC4 in Figure \ref{f2}, in which
the BML region of SC4 (the red contours) is distinct from
the quiescent gas traced by \thCO\ emission (the blue contours).
Especially, the shocked molecular gas shells traced by \twCO\
($J$=3--2) emission (the red contours in Figure \ref{f2}) seem to 
resemble the unperturbed molecular shells traced by \thCO\ 
emission (the thick-dashed gold lines in Figure \ref{f2}).
These features imply that the shocked gas is indeed physically
associated with the quiescent MCs at $V_{\rm LSR}\sim +60\km\ps$.

We try to investigate the distribution of the shocked 
gas in the whole region of HESS J1912$+$101.
However, the \twCO\ ($J$=1--0) line profile may be contaminated by 
unrelated MCs along the same line of sight (LOS) in the direction of 
the inner Milky Way. Observations in higher CO transitions might 
show less confusion because the broadline emission in high $J$ is 
relatively strong with respect to the quiescent MCs. 
Unfortunately, the whole region of HESS J1912$+$101 is not 
completely covered by the COHRS \twCO\ ($J$=3--2) survey in the 
direction (see Table 1). 

We attempt to construct a map of 
$I$(\twCO , $v1:v2$) $-$ $I$(\thCO , $v1:v2$)$\times7$
to illustrate the possible distribution of the shocked material,
where $I$(\twCO , $v1:v2$) and $I$(\thCO , $v1:v2$) 
are the integrated intensity of \twCO\ ($J$=1--0) and \thCO\ ($J$=1--0) 
in the velocity range of $v1$ to $v2$, respectively.
Here, we adopt $v1$=+67~$\km\ps$ and $v2$=+75~$\km\ps$. It roughly represents the
velocity range of the possible redshifted CO wing component.
The map will reveal the \twCO\ ($J$=1--0) enhancement 
($\gsim 6$~K~$\km\ps$) in the velocity range of 67--75~$\km\ps$, which
may sketch the distribution of the shocked gas after subtracting
the contribution of the unrelated CO emission (controlled
by $I$(\thCO , $v1:v2$)$\times7$) in the whole FOV toward
HESS J1912$+$101.

The aim of the above operation is to reveal the potential redshifted-BML 
regions based on the \twCO\ and \thCO\ ($J$=1--0) lines. Generally, the emission
of typical MCs is optically thick for \twCO\ line and optically thin
for \thCO\ line. On the contrary, the shocked molecular gas will exhibit 
optically thin \twCO\ emission with little \thCO\ emission for the broadened
wing component. 
Using $I$(\twCO , 67:75) $-$ $I$(\thCO , 67:75)$\times7$, we
can filter out much of the molecular emission from the unrelated MCs 
in 67--75~$\km\ps$, leaving the enhanced \twCO\ emission in the 
same velocity interval along the LOS.
The revealed \twCO\ enhancement is probably 
related to the BML regions for shocked gas. The shortcoming 
is that some shocked features in the map are lost because of the confusion 
between the unrelated MCs and the shocked gas along the LOS (e.g., SC1).
Some realistic BML regions cannot be identified either due to
the weak emission of the broadened \twCO\ wing and the high 
criterion of $\Delta I \gsim 6$~K~$\km\ps$ (e.g., SC3).

\begin{figure}
\plotone{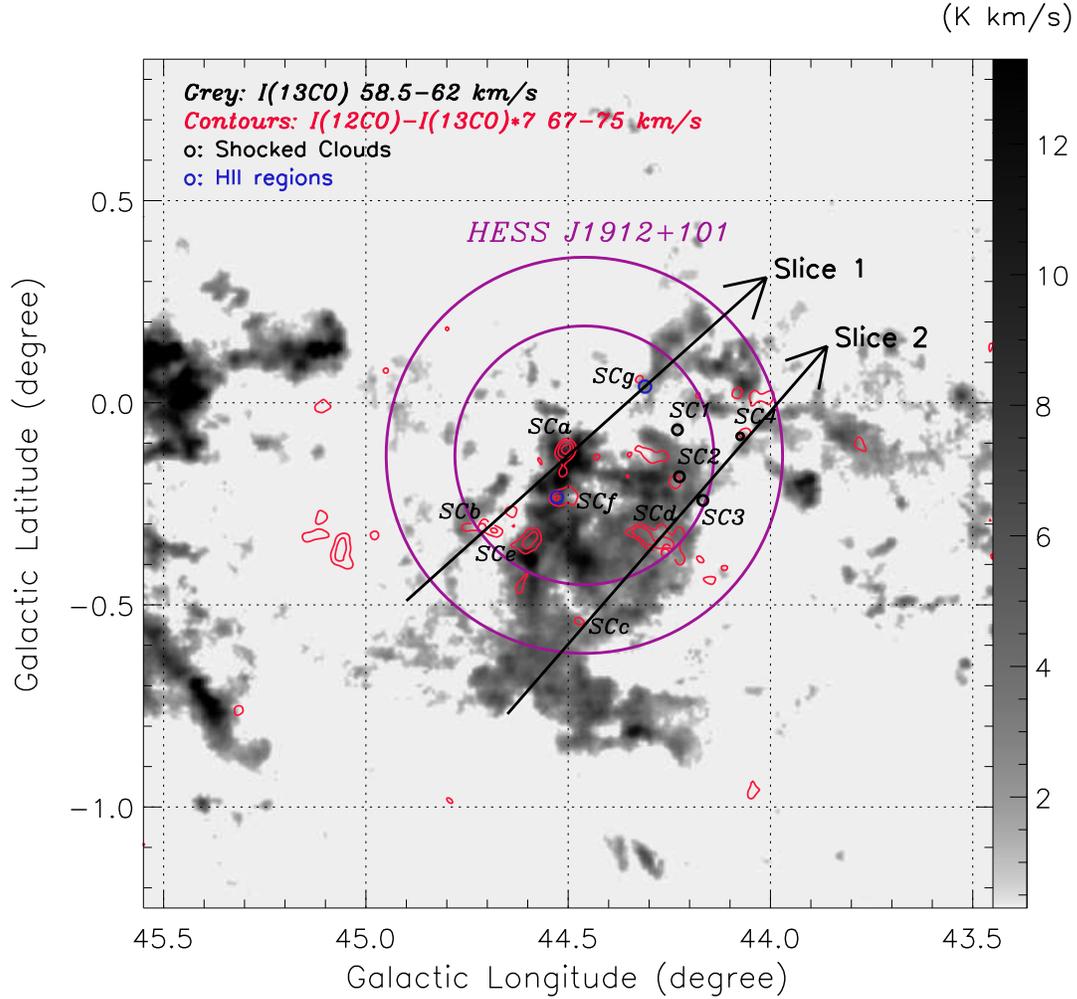}
\caption{Integrated \thCO\ ($J$=1--0) emission toward HESS J1912$+$101
in the interval of 58.5--62.0~km~s$^{-1}$, overlaid with the red contours
(6, 8, and 10~K$\km\ps$) of possible \twCO\ ($J$=1--0) wing component in the 
interval of 67--75~km~s$^{-1}$ (see text). 
The two purple circles indicate the outer and inner TeV
shells of HESS J1912$+$101.
The two arrows of Slice 1 and Slice 2 indicate PV
slices shown in Figure \ref{f5}. Positions and sizes of shocked BML regions
of SC1--SC4 (see spectra in Figure \ref{f3}) are also labeled.
SCa--SCe indicate the possible shocked BML regions revealed from the
\twCO\ ($J$=1--0) wing component, while SCf and SCg may be 
contaminated by the nearby \HII\ regions (the two blue circles for 
HRDS G044.528$-$0.234 and VLA G044.3103$+$00.0410; see the text).
\label{f4}}
\end{figure}

\begin{figure}
\plottwo{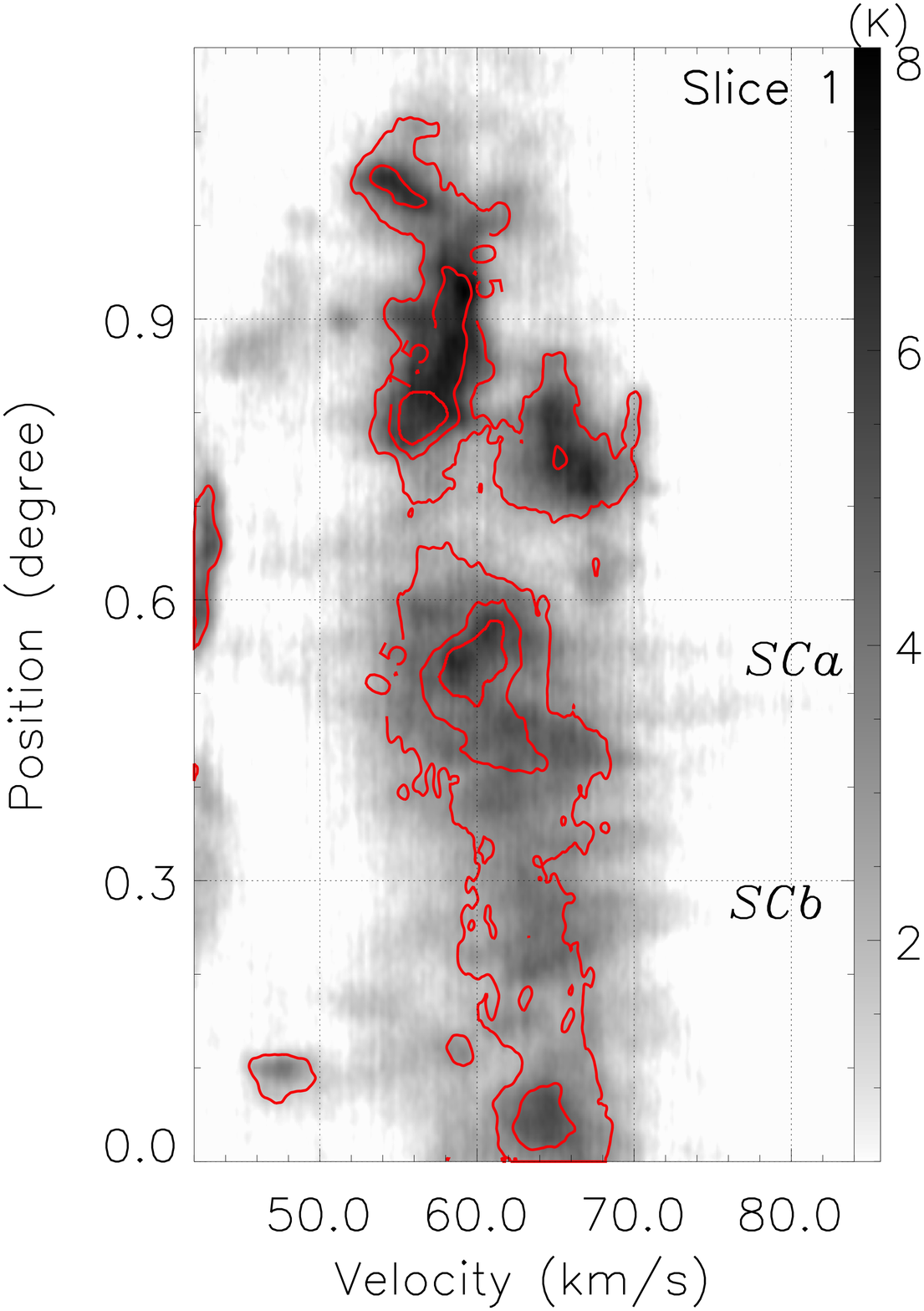}{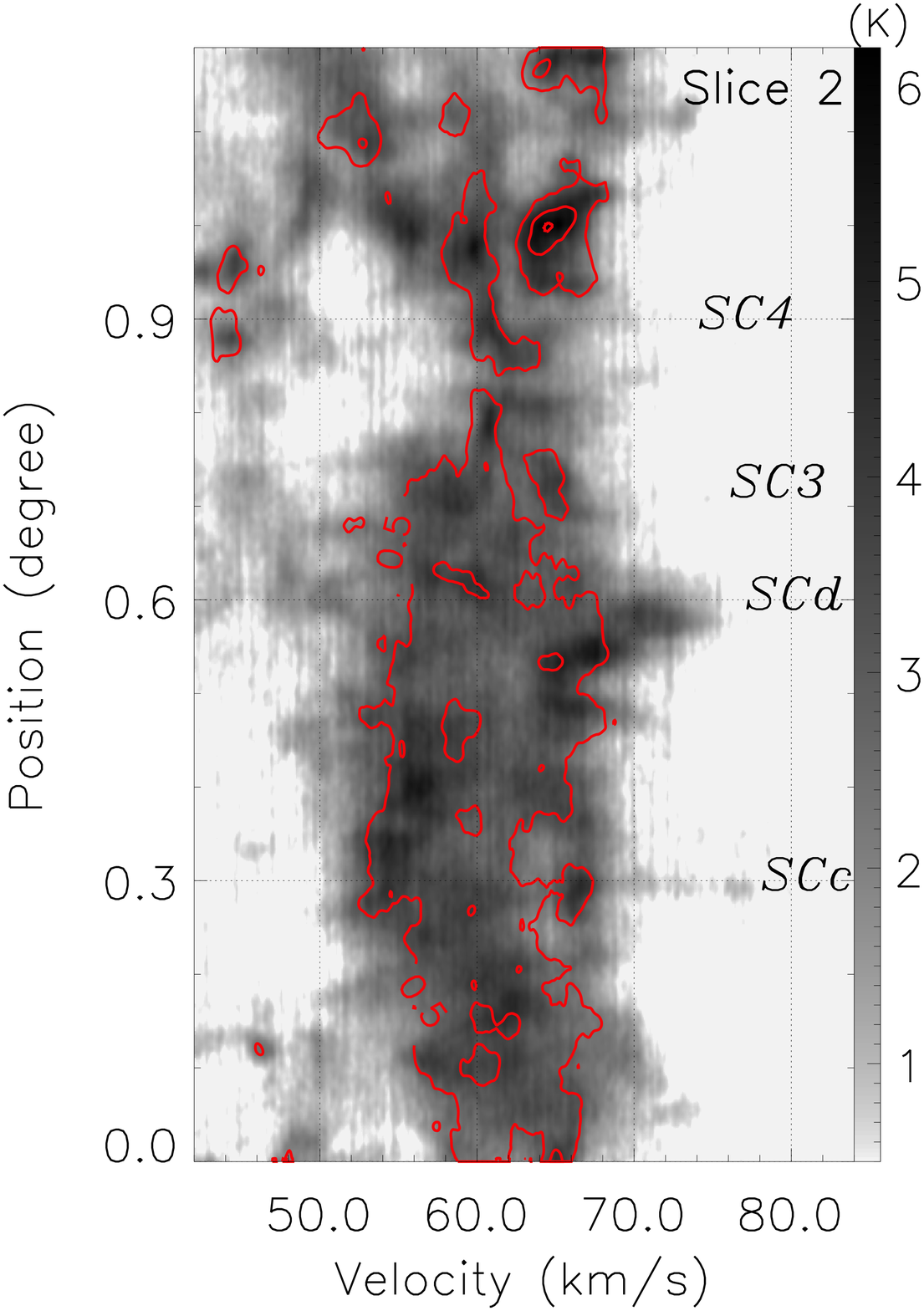}
\caption{Left panel: PV diagram of \twCO\ ($J$=1--0) emission along Slice 1,
overlaid with the contours of \thCO\ emission. The overlaid contour
levels start from 0.5 K and increase by a step of 1 K.
The PV slice has a length of 1\fdg2
(from ($l=$44\fdg90, $b=-$0\fdg49) to ($l=$44\fdg01, $b=$0\fdg31))
and a width of 5\farcm5 (see the arrow Slice 1 in Figure \ref{f4}).
Right panel: PV diagram of \twCO\ ($J$=1--0) emission along Slice 2,
overlaid with the contours of \thCO\ emission. The overlaid contour
levels also start from 0.5 K and increase by a step of 1 K.
The PV slice has a length of 1\fdg2
(from ($l=$44\fdg65, $b=-$0\fdg77) to ($l=$43\fdg86, $b=$0\fdg14))
and a width of 5\farcm5 (see the arrow Slice 2 in Figure \ref{f4}).
\label{f5}}
\end{figure}

\begin{figure}
\plotone{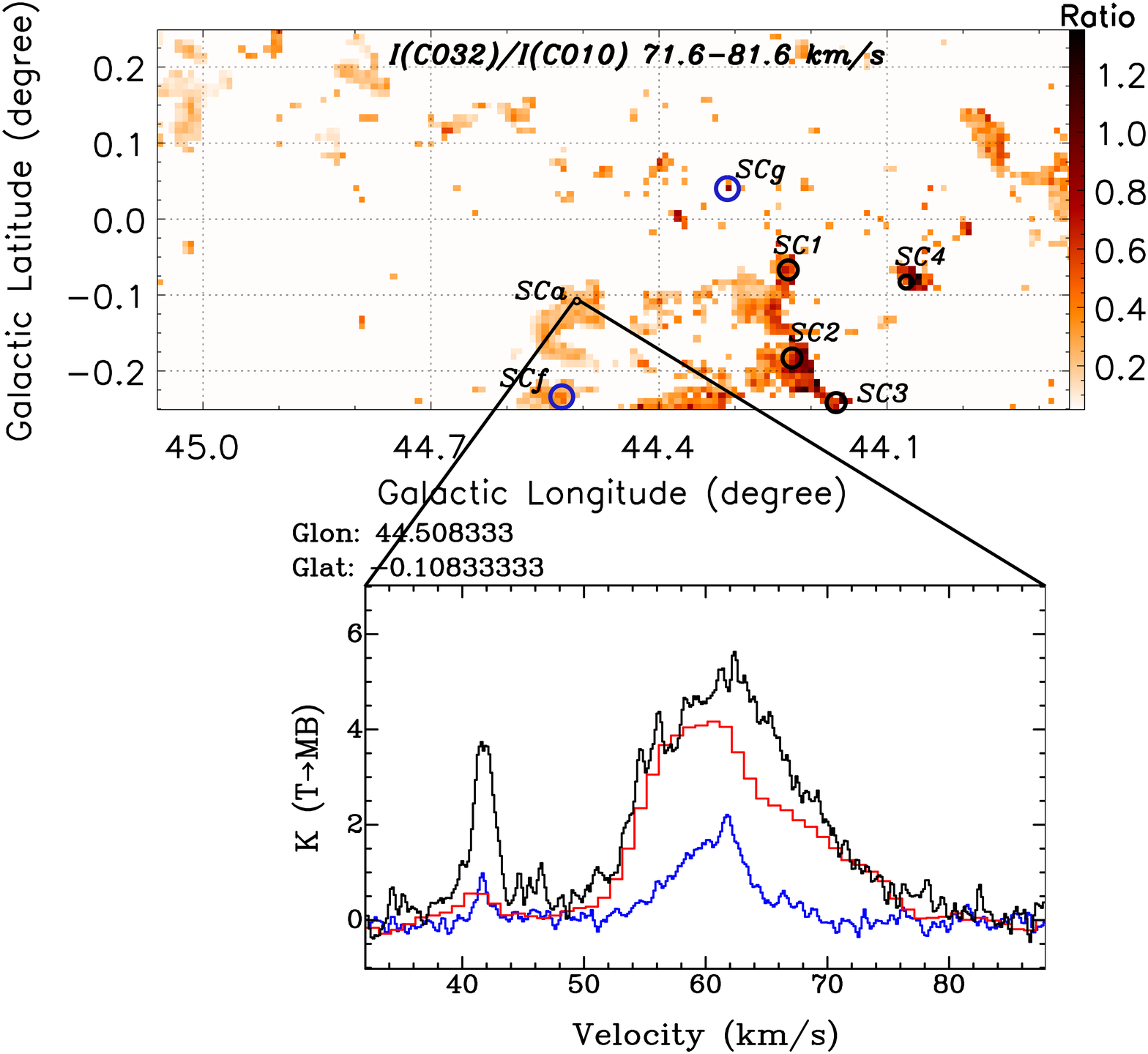}
\caption{Map of the \twCO\ $J$=3--2/1--0 intensity ratio toward HESS J1912$+$101
at the velocity range of 71.6--81.6~km~s$^{-1}$ after convolving the COHRS 
\twCO\ ($J$=3--2) data to the same beam size as the MWISP \twCO\ ($J$=1--0)
data. The spectra of position $l$=44\fdg508 and $b$=$-$0\fdg108 with size of
0\farcm5 in SCa are also given, where \twCO\ ($J$=1--0), \thCO\ ($J$=1--0),
and \twCO\ ($J$=3--2)$\times$2 are in black, blue, and red, respectively.
Note that the unconvolved \twCO\ ($J$=3--2) spectrum is shown to
highlight the redshifted-broadening. 
The two blue circles show the \HII\ regions of
HRDS G044.528$-$0.234 and VLA G044.3103$+$00.0410. 
\label{f6}}
\end{figure}

Figure \ref{f4} displays the distribution of the possible 
BML regions, in which the enhanced \twCO\ ($J$=1--0)
emission in 67--75~$\km\ps$ (the red contours) is overlaid with the 
\thCO\ emission in the velocity interval of 58.5--62.0~$\km\ps$.
An intriguing result is that 
most of the \twCO\ ($J$=1--0) enhancements
are concentrated within the TeV shell. We make two PV diagrams 
along Slice 1 and Slice 2 (see Figure \ref{f4}) to search for 
possible kinematic features of the shocked gas further.
In the PV diagrams (Figure \ref{f5}), 
the redshifted protrusions of SCa--SCd in 
\twCO\ ($J$=1--0) emission are actually discerned from the quiescent \thCO\ 
emission, suggesting that the samples of SCb--SCe
are likely BML regions.
We note that shocked molecular gas at SC2 and SC4 from 
the previous \twCO\ ($J$=3--2) analysis is coincident well with 
its nearby strong \twCO\ ($J$=1--0) wing emission in the velocity
range of 67--75$\km\ps$. 
Moreover, the identification of the possible shocked BML region SCa at 
($l$=44\fdg508, $b=-$0\fdg108) is also confirmed by its broadened 
\twCO\ ($J$=3--2) emission up to $V_{\rm LSR}\sim +77\km\ps$
(see the extracted spectrum in Figure \ref{f6}), 
strengthening the validity of this method.

Some BML regions identified from the \twCO\ ($J$=3--2) emission
(e.g., SC1 and SC3) do not correspond to red contours,
indicating the missing identification in Figure \ref{f4}.
Of course, some fragmental structures of red contours outside 
the TeV shell are not the BML region either. 
Additionally, \HII\ regions of HRDS G044.528$-$0.234
\citep{2011ApJS..194...32A} and VLA G044.3103$+$00.0410
\citep{2009A&A...501..539U} are associated with the strong
\twCO\ ($J$=1--0) emission in 67--75~$\km\ps$ of SCf and SCg, respectively.
All of these show the complexity for the 67--75$\km\ps$ 
CO emission in the large-scale FOV. Therefore, the red contours in 
Figure \ref{f4} only roughly outline the distribution of the 
possible BML regions in the field (e.g., SCa--SCe).

Further observations and analysis are needed 
in order to identify the more reliable distribution of the shocked 
molecular gas in the whole region.
As seen in Figure \ref{f6}, the intensity ratio between 
\twCO\ ($J$=3--2) emission and \twCO\ ($J$=1--0) emission is high for 
BML regions in the velocity interval of 71.6--81.6$\km\ps$ 
(or the redshifted-broadening part), which is consistent with the 
suggestion that an enhanced high-to-low ratio in CO line wings 
of MCs is a clear and useful indicator of shocked gas
\citep[e.g.,][]{1998ApJ...505..286S}.
High $J$ observations of CO are thus helpful in providing 
the unambiguous, direct signature of the distribution of 
the shocked material in the whole region of the TeV source.

\subsection{High-velocity \mbox{H\,\textsc{i}} gas}
In the above section, we have shown that the emission of the shocked 
molecular gas, which exhibits significant redshifted-broadening 
relative to the ambient gas, is concentrated toward the shell-like
TeV source of HESS J1912$+$101 (Figures \ref{f2}, \ref{f4}, and \ref{f6}). 
The LSR velocity of the shocked 
molecular gas exceeds that of the tangent points in the direction 
($l$=44\fdg5, $v_{\rm tangent}$= +71~$\km\ps$). As a result, 
one can readily discern that the disturbed gas from the surrounding emission 
if its LSR velocity is higher enough than the tangent point's velocity.
Accordingly, we wonder whether the possible high-velocity atomic gas 
can be discerned from the background emission 
in the FOV.

Although the \mbox{H\,\textsc{i}} emission displays a complex 
spatial morphology and is more diffuse than 
CO emission in the FOV, two features of the high-velocity atomic gas 
are actually revealed toward the concentration of the shocked molecular gas
based on VGPS \mbox{H\,\textsc{i}} data (see the blue contours in Figure \ref{f2}).
The LSR velocity of the atomic gas 
is high enough so that such expanding features can be distinguished
from the surrounding background emission at positive velocities
of 97.5--111.5$\km\ps$. The high-velocity atomic gas, which is 
observed projected close to the western edge of the TeV source, is
spatially coincident with the surrounding molecular partial shell 
structures 
at $\sim +60\km\ps$ traced by \thCO\ emission 
(see the thick-dashed gold lines in Figure \ref{f2}).

The atomic gas emission displays partial shell structures
from northeast to southwest
(length$\gsim 20'$ and width$\sim 4'$; see the
blue contours and the thick-dashed blue lines in Figure \ref{f2}).
The emission of the high-velocity atomic gas is very faint ($\lsim $2--3~K),
but the shell structures are distinct from the
background emission in the integrated intensity map
of $V_{\rm LSR} =97.5-111.5\km\ps$.
The \mbox{H\,\textsc{i}} emission 
of the shell structures drops dramatically at $V_{\rm LSR}\sim +100\km\ps$
and seems to disappear at $V_{\rm LSR}\sim +110\km\ps$.
We cannot discern any \mbox{H\,\textsc{i}} structures
from the integrated map of $V_{\rm LSR}\ge +112\km\ps$.
At velocities $\sim$80--90~$\km\ps$, the
brightness temperature of the \mbox{H\,\textsc{i}} shells
is $\sim$10~K, but the shell features are only marginally discerned
due to confusion from the close by diffuse background emission.

The high-velocity \mbox{H\,\textsc{i}} gas Shell-a seems to be 
situated slightly outside the $\sim+60\km\ps$ molecular shell
(Figure \ref{f2}).
It is also interesting to note that the shocked molecular gas 
(red contours for \twCO~($J$=3--2) emission), the quiescent molecular
gas (gray image for \thCO~($J$=1--0) emission), and the high-velocity
atomic gas (blue contours for \mbox{H\,\textsc{i}} emission) 
are in contact with each other and 
seem to display an ordered arrangement from the inner side 
to the outer side (or from southeast to northwest),
indicating that these shells are probably physically associated.
The close morphological agreement between the shocked molecular
gas and the high-velocity \mbox{H\,\textsc{i}} gas, together with
the redshifted velocities of the disturbed gas with respect to 
the +60$\km\ps$ quiescent gas, indicates that the gas shells probably
have the same origin (see discussions in Section 3.3).

We summarize the properties of the high-velocity \mbox{H\,\textsc{i}} shells
in Table~2. It should be noted that the estimated column density in Table~2
should be regarded as lower limits since the disturbed atomic gas at 
low velocities (e.g., 60--100$\km\ps$) may has relatively higher $T_{\rm mean}$ 
than our estimates of $\sim$1~K at +100$\km\ps$. Accordingly, 
the total mass and kinetic energy of the disturbed neutral gas in
the Shell-a and Shell-b should also be viewed as lower limits.

\subsection{An old SNR: its distance, age, and progenitor}
So far, about one fourth of SNRs are thought to be associated with 
MCs \citep[refer to, e.g.,][]{2010ApJ...712.1147J,2014IAUS..296..170C}. 
These SNR--MC systems often display 
broadenings, wings, or asymmetric profiles in the molecular
line spectra. On the other hand, some systematic searches for 
\mbox{H\,\textsc{i}} shells associated with Galactic SNRs were 
performed \citep[e.g.,][]{1991ApJ...382..204K,2004JKAS...37...61K}.
Several expanding \mbox{H\,\textsc{i}} shells are also confirmed
to be associated with the SNR's radio emission
\citep[e.g.,][]{2013ApJ...777...14P}. 

These studies suggest that a significant fraction of the SNR's 
mechanical energy is stored in the ambient interstellar medium 
(ISM). Disturbed by the SNR shock, 
the material in the vicinity of the remnant will be swept up
and exhibit expanding molecular and/or atomic shell-like structures 
on a large scale. These shell-like features are often
incomplete because of the inhomogeneous ISM.

Both shocked molecular gas and the associated expanding
\mbox{H\,\textsc{i}} shells are found to be physically associated with
MCs at $V_{\rm LSR}\sim +60\km\ps$
(see Sections 3.1 and 3.2).
It is thus natural to ascribe the disturbed gas to
the perturbation of an SNR in the region.
In such a scenario, the SNR provides the required kinetic energy and
momentum for its ambient disturbed material.
Broadly speaking, \mbox{H\,\textsc{i}} gas, tracing emission of
more diffuse gas, may be swept farther away than molecular gas
if the interaction happens early enough. 
As seen in Figure \ref{f2}, Gas Shell-a may be the expected case 
that the high-velocity atomic gas displays more redshifted velocity 
structure with respect to that of the shocked molecular gas 
and expands somewhat farther toward the northwestern direction than the 
associated CO shell.

The separation between the molecular shell and the 
atomic shell is about 2' (or 2.4~pc at a distance of 4.1~kpc,
see Gas Shell-a in Figure \ref{f2}). 
If the SNR encounters the molecular shell, the shock is significantly decelerated
by the interaction with the dense gas. MCs are known to be clumpy 
\citep[e.g., a volume filling factor of $< 0.1$ for dense clumps,][]{1993prpl.conf..125B}. 
That is, the SNR shock
evolves mainly in the interclump medium (e.g., $n_{\rm H}\sim$10 H atoms
cm$^{-3}$). Therefore, the shock in the interclump medium and the surrounding
ISM with lower
density may be still moving while the shock in dense molecular clumps
nearly stalls out. The dynamic time of the moving atomic gas thus can be estimated
as $\Delta R$/$v_{\rm exp} \sim$2.4~pc/40~$\km\ps$=0.6$\times10^{5}$~years.
The dynamic timescale of the moving atomic shell is
slightly smaller and comparable with the remnant's age of $\gsim 0.7\times10^5$~years
(see below). On the other hand,
both of the atomic and the molecular shells seem to almost overlap
at the outer region (or the northwestern region, 
see Gas Shell-b in Figure \ref{f2}), suggesting the recent interaction 
between the shock and the dense gas at the boundary of the remnant.

One might wonder whether the shocked molecular gas, as well as 
the associated high-velocity \mbox{H\,\textsc{i}} shells, 
is related to other energetic sources, such as nearby \HII\ regions 
and/or the stellar winds from massive OB stars. However, these 
possibilities can be excluded because of lacking the associated 
energetic sources in the direction. Some \HII\ regions with small 
sizes of $\lsim 2'$ \citep[e.g.,][]{2011ApJS..194...32A,2014ApJS..212....1A}
are distributed in the FOV, but they are at least $\sim6'$ 
(7~pc at a distance of 4.1~kpc) away from identified 
shocked BML regions. Actually, the overlapping case 
between \HII\ regions and BML regions can be exactly picked out because 
\HII\ regions have bright IR and non-thermal radio emission
(see the two blue circles in Figures \ref{f4} and \ref{f6}).
A young massive cluster of Mercer 20 located at 
($l$=44\fdg16, $b$=$-$0\fdg07) 
contains several massive stars within a radius of $\sim1'$
\citep{2005ApJ...635..560M,2009ApJ...697..701M}. 
Although these OB stars at a distance of 
$\sim3.8$~kpc \citep{2009ApJ...697..701M} or $\sim8.2$~kpc 
\citep{2015A&A...575A..10D} are projected on the northeastern border 
of the high-velocity \mbox{H\,\textsc{i}} shells, they are 
unlikely responsible for the large-scale neutral atomic shells 
with sizes of $\sim$0\fdg4 (or 28.6~pc at 4.1~kpc) and 
the widely distributed BML regions. 

We thus suggest that an old SNR at a distance 
of 4.1~kpc (see below) appears to be a promising candidate for 
the surrounding shocked molecular gas and the high-velocity
neutral atomic shells, although no corresponding radio feature
has been detected in the VGPS 1.4 GHz continuum emission.
The missing radio structure is probably due to the faint 
synchrotron emission of the old SNR (see Section 3.4),
the sensitivity limitation of the radio observation,
the contamination of the diffuse radio emission in the Galactic plane,
and the confusion of bright radio sources projected onto the area.
Further high-resolution and sensitivity radio observations
toward the region should be carried out to investigate the
possible non-thermal radio emission.

In subsequent analysis, we investigate properties of 
the likely SNR. First, the distance is one of the most important 
values for studying the nature of the source. Spectroscopic information 
of the quiescent molecular gas is thus helpful
for identifying the LSR velocity of MCs and associating its kinematic
distance to the target that we are interested in.
For the $V_{\rm LSR}$=+60~$\km\ps$ molecular gas, we find that most of 
\thCO\ peaks from the shocked gas and the molecular shells
correspond to \mbox{H\,\textsc{i}} dips, indicating that they 
are likely at near distance based on the \mbox{H\,\textsc{i}}
self-absorption method \citep{2009ApJ...699.1153R}.
Accordingly, we adopt a near distance as the distance of the +60$\km\ps$ 
GMC and exclude the far distance.
Using the A5 rotation curve model of \cite{2014ApJ...783..130R},
we place the +60$\km\ps$ GMC and the associated SNR at
a near kinematic distance of 4.1 kpc.

Second, as the brightness temperature of the \mbox{H\,\textsc{i}} 
shells is relatively low ($\sim$1~K at +100$\km\ps$), we may assume that
the high-velocity atomic gas is optically thin. Then the \mbox{H\,\textsc{i}} 
mass of the high-velocity shells is estimated to be 
$\gsim230d_{4.1}^{2}$~$M_{\odot}$, where $d_{4.1}$ is the distance to 
the cloud in units of 4.1~kpc. Adopting a radius of 0\fdg405 
\citep[e.g., (0\fdg49+0\fdg32)/2,][]{2016arXiv161200261G}, 
the solid angle of the high-velocity 
\mbox{H\,\textsc{i}} shells is about 0.24~sr. 
The derived total \mbox{H\,\textsc{i}} mass for a sphere with a radius 
of 0\fdg405 is $\gsim1.2d_{4.1}^{2}\times10^4$~$M_{\odot}$, which 
corresponds to a kinetic energy of $\gsim1.9d_{4.1}^{2}\times10^{50}$~ergs
for the expansion velocity of $v_{\rm exp}\gsim40\km\ps$.
A small fraction of 10$\%$ energy from a single supernova 
explosion with a total mechanical energy of 1.9$\times10^{51}$~erg
is sufficient to power the expanding atomic gas, favoring
an SNR origin for the disturbed gas that is widely distributed 
in the FOV.

Third, the pre-explosion ambient density is estimated to be
$\sim3.7d_{4.1}^{-1}$~cm$^{-3}$ if the 
neutral gas with total mass of $1.2d_{4.1}^{2}\times10^4$~$M_{\odot}$
was uniformly distributed within a sphere of a radius of 29.0 pc
(0\fdg405 at a distance of 4.1 kpc).
According to \cite{1988ApJ...334..252C} for an
SNR in the pressure-driven snowplow phase, the explosion
energy could be: 
$E_{\rm SN}=6.8\times10^{43} n_{0}^{1.16} v_{\rm exp}^{1.35} R_{\rm s}^{3.16} 
\zeta_{\rm m}^{0.161}$~erg 
$\sim1.9d_{4.1}^{2}\times10^{51}$~erg, where $n_{0}=3.7d_{4.1}^{-1}$~cm$^{-3}$, 
$v_{\rm exp}=40\km\ps$, $R_{\rm s}=29.0d_{4.1}$~pc, 
and $\zeta_{\rm m}=Z$/$Z_{\odot}=1$.
The estimated explosion energy of $1.9d_{4.1}^{2}\times10^{51}$~erg
is slightly larger than the typical value of $1\times10^{51}$~ergs,
but still reasonable. 
If the material inside the volume is not uniformly
distributed, the above total mass of $1.2d_{4.1}^{2}\times10^4$~$M_{\odot}$ 
is probably overestimated, leading to the larger explosion energy
for an SNR. It is a more realistic case because the neutral atomic shells 
with the highest velocity of 97.5--111.5$\km\ps$ are only detected near 
the western boundary of HESS J1912$+$101.

The age of the SNR could be 
$2R_{\rm s}$/($7v_{\rm exp})\lsim$2.0$\times10^{5}$~years
for a radius of 29.0~pc and an expansion velocity of $v_{\rm exp}\gsim40\km\ps$
in the radiative pressure-driven snowplow phase.
We stress that the above value is an upper limit 
due to the somewhat underestimated velocity of the remnant's shock.
Some atomic gas structures with higher velocity but lower
intensity may be lost because of confusion from the 
diffuse emission of the Galactic plane.
Additionally, the evolutionary phase of the remnant should be speeded up 
if the SNR evolves in the environment of a pre-existent wind bubble
\citep[e.g.,][]{2005ApJ...630..892D,2007ApJ...667..226D}.

The lower limit of the SNR age may be obtained from 
the view of radiative cooling of the remnant. 
According to little X-ray emission and the low shock velocity
of $\sim40\km\ps$ in our case, we suggest that the SNR has
lost a substantial amount of explosion energy from radiative cooling. 
The forward shock of the remnant is also significantly decelerated. 
The SNR age is thus at least 2--3 times larger than the transition time, 
which is defined as the time of an adiabatic Sedov--Taylor blast wave to a radiative 
pressure-driven snowplow phase. The transition time is given by
$t_{\rm {tr}}\approx2.9\times10^{4} (\frac{E_{\rm SN}}{{\rm 10^{51}~erg}})^{4/17} 
(\frac{n}{{\rm 1~cm^{-3}}})^{-9/17}$~years \citep[e.g.,][]{1998ApJ...500..342B}.
Assuming $n\sim{\rm 1~cm^{-3}}$, $E_{\rm SN}\sim{\rm 1\times10^{51}~erg}$,
and $t_{\rm SNR}\gsim3\times t_{\rm tr}$, the SNR age is $\gsim0.9\times10^5$~years, 
which is reasonably consistent with numerical simulations of 
$\gsim$(0.7--1.1)$\times10^{5}$~years 
that the SNR has radiated most of its explosion energy in the wind-bubble environment
\citep[e.g., Figures 3 and 8 in][]{1991MNRAS.251..318T}.
The age of the SNR is thus (0.7--2.0)$\times10^{5}$~years.

We also note that two pulsars, PSR J1913$+$1000 and PSR J1913$+$1011,
are situated within the inner shell of HESS J1912$+$101. PSR J1913$+$1000
\citep{2004MNRAS.352.1439H} is not related to
the old SNR because of their different distances
($d_{\rm PSR}\sim$7.9~kpc vs. $d_{\rm SNR}\sim$4.1~kpc)
and ages ($t_{\rm PSR}\sim$7.9$\times10^{5}$~years vs.
$t_{\rm SNR}\sim$(0.7--2.0)$\times10^{5}$~years).
Meanwhile, PSR J1913$+$1011, which is located at ($l$=44\fdg48, $b=-$0\fdg17)
and is close to the geometric center of HESS J1912$+$101 (see Figure \ref{f1}), 
has a distance of 4.5~kpc
from dispersion measurements and an age of 1.7$\times10^{5}$~years
from the spin-down method.
Both of the two values are consistent well with the estimated
distance and age of the remnant, indicating that PSR J1913$+$1011 
is probably the result of the core collapse of the massive
progenitor of the old SNR.

The progenitor's wind of the old SNR may excavate its surrounding and
form a pre-existing wind-blown bubble before the supernova explosion.
Assuming that the remnant's progenitor was a massive single star and
the current molecular shells are mainly generated by its massive wind,
we can estimate its mass from the cavity size.
The size of the $+60\km\ps$ GMC is comparable to the TeV shell's size
of HESS J1912$+$101 (Figures \ref{f1} and \ref{f4}).
Adopting the inner radius of 22.9 pc (0\fdg32) as the radius of the wind bubble,
an initial stellar mass of $\gsim$26~$M_{\odot}$ is obtained from the 
linear relation between the progenitor's mass and the wind bubble's radius
\citep[see Equation 8 and Figure 1 in][]{2013ApJ...769L..16C}.
It implies that the progenitor's spectral type is probably
earlier than O9V in a constant interclump pressure of
$p/k\sim 10^{5}$~cm$^{-3}$~K.

Briefly, an old SNR seems to be the origin of the shocked molecular gas
and high-velocity \mbox{H\,\textsc{i}} shells.
In this case, the mass of the ambient gas is swept up by the wind
of the massive progenitor of the SNR and the gas in the formed shell 
is accelerated by the SNR's shock. The kinetic energy of the disturbed 
gas is also provided by the remnant, resulting in the 
remarkable morphological and kinematic features of the molecular and
atomic gas shown by us (Figure \ref{f2}).
The non-detection in the X-ray bands, together with the absence of 
corresponding radio structures in the surrounding, is consistent with 
the SNR's old age of (0.7--2.0)$\times10^{5}$~years (see Section 3.4).

\subsection{Origin of the shell-like TeV source of HESS J1912$+$101}
The PWN and the SNR scenarios for HESS J1912$+$101 were
discussed recently \citep{2008AA...484..435A,2015ICRC...34..886P}.
However, the origin of the TeV source is yet unconfirmed.
It is worth mentioning that TeV sources of ARGO J1912$+$1026 
\citep[centered at $l=$44\fdg59 and $b=$0\fdg20, the angular 
resolution $\sim$0\fdg17--1\fdg66,][]{2013ApJ...779...27B} and 
2HWC J1912$+$099 \citep[centered at $l=$44\fdg15 and $b=-$0\fdg08,
the angular resolution $\sim$0\fdg2--1\fdg0,][]{2017ApJ...843...40A}
were proposed to be associated with HESS J1912$+$101. All of these 
characteristics
indicate that the high-energy TeV emission is significantly enhanced
in such a region. 
In our opinion, the high-energy $\gamma$-ray emission in the region
can be explained by interactions between accelerated particles from 
the SNR and the surrounding dense medium. 

We have revealed that shocked MCs, together with the associated 
high-velocity atomic gas, are concentrated toward HESS J1912$+$101.
Especially, the shocked MCs with broadline wings, as well as the 
expanding \mbox{H\,\textsc{i}} shells (Figure \ref{f2}), 
are observed mainly projected close to the western boundary of 
the TeV source. These interesting features indicate large amounts 
of momentum and mechanical energy injection to the disturbed gas.  
We suggest that 
an energetic source is initially required at the 
southeast region
of the gas shells to account for the kinematic features of the 
disturbed gas on a large scale. An old SNR 
may be a good choice and a reasonable candidate because of 
lacking other energetic sources, e.g., 
giant \HII\ regions or massive OB stars (see Section 3.3). 
As a consequence, the SNR would provide the required energy of the 
high-energy protons.

The angular resolution of H.E.S.S. observation is better than
0\fdg1, which allows us to compare the TeV emission with the distribution
of the shocked and/or dense molecular gas with the precision of
several arcmin.
The distribution of the high-energy emission of HESS J1912$+$101 
is not uniform. Several enhanced TeV peaks ($\gsim 5\sigma$) can be 
seen at ($l\sim$~44\fdg3, $b\sim$~0\fdg1), 
($l\sim$~44\fdg2, $b\sim-$0\fdg3), ($l\sim$~44\fdg6, $b\sim-$0\fdg4), 
($l\sim$~44\fdg8, $b\sim-$0\fdg3), and ($l\sim$~44\fdg7, $b\sim$~0\fdg0).
Based on CO data, we find that the shocked gas and/or dense MCs are
spatially coincident with the VHE $\gamma$-ray emission. 

The size of the $60\km\ps$ MC concentration is comparable to
that of the TeV source of HESS J1912$+$101 (Figures \ref{f1}
and \ref{f4}).
Generally, the overall \thCO\ emission is strong 
in the western part of the TeV shell, which is consistent with the 
brighter high-energy emission on the western half of HESS J1912$+$101.
TeV source 2HWC J1912$+$099 
\citep{2017ApJ...843...40A} appears projected onto such the region,
in which the disturbed ISM is also concentrated
(see Figure \ref{f2}). The BML regions (e.g., SCa, SCb, SCd, SCe, 
and SC1--SC4) show good spatial coincidences with the TeV 
enhancement (Figure \ref{f4}). At ($l\sim$~44\fdg3, $b\sim$~0\fdg1), 
the dense (traced by \thCO\ emission, $\sim10^2$--$10^3$~cm$^{-3}$) 
and massive ($\sim3\times10^4$~$M_{\odot}$) MC complex at $\sim$+60$\km\ps$, 
which is related to an IR bubble N91 \citep{2006ApJ...649..759C}
at a massive star-forming region \citep[e.g., Gong et al. 2017 
(to be submitted) and][]{2012AJ....144..173D},
is spatially coincident with the brightest TeV emission there.
On the contrary, the eastern \thCO\ emission is relatively weak.
However, a piece of \thCO\ cloud (size$\sim3'$ and $V_{\rm LSR}\sim +58\km\ps$),
together with more extended \twCO\ emission,
is located at ($l\sim$~44\fdg72, $b\sim$0\fdg03). Some patches of molecular
gas with weak \thCO\ emission also can be seen along the eastern 
inner shell of the TeV source (Figures \ref{f1} and \ref{f4}), 
which may be related to the relatively weak 
$\gamma$-ray emission on the eastern half of HESS J1912$+$101.

The detection of the \mbox{H\,\textsc{i}} and CO partial shell-like 
structures, as well as the C$^{18}$O emission in some regions, shows 
that the old SNR is located in the complex and dense ISM.
Naturally, the established source-target (or SNR--MCs) system favors 
the hadronic scenario for the observed TeV emission
\citep[e.g.,][]{2006MNRAS.371.1975Y,2009MNRAS.396.1629G}. 
The hadronic $\gamma$-ray flux can be estimated by \cite{1994A&A...287..959D}
as $F_{\rm \gamma} (>{\rm 0.681~TeV}) \approx 9\times10^{-11} \Theta (E/{\rm 1~TeV})^{-1.1}
(E_{\rm SN}/{\rm 10^{51}~erg}) (d/{\rm 1~kpc})^{-2} (n/{\rm 1~cm^{-3}})$. 
Here, $\Theta$ is the fraction of the total supernova explosion energy 
converted to CR. Using numerical values of 
$F_{\rm \gamma} (> {\rm 0.681~TeV})$=4.5$\times10^{-12}$~photons~cm$^{-2}$~s$^{-1}$,
$\Theta$=0.1--0.01, $E_{\rm SN}=10^{51}$~erg, and $d$=4.1~kpc, 
the hadronic scenario requires a mean target gas density of 6--60~cm$^{-3}$
for the surrounding ISM, which is consistent with our 
finding of the molecular environment there.

According to the spectral fit of HESS J1912$+$101 \citep{2016arXiv161200261G},
the energy flux in the 1--10~TeV band is estimated to be
$F(\rm 1-10~TeV)$=8.1$\times10^{-12}$~erg~cm$^{-2}$~s$^{-1}$.
Using $Chandra$ data, \cite{2008ApJ...682.1177C} found that the 
upper limit of the X-ray flux in the 0.3--10~keV band is 
0.7$\times10^{-13}$~erg~cm$^{-2}$~s$^{-1}$ for the center of gravity of
HESS J1912$+$101. It indicates that the flux ratio of 
$R_{\rm TeV(1-10~TeV)/X-ray(0.3-10~keV)}$ should be larger than 120.
%because of the relatively strong absorption in the soft X-ray band. 
The high flux ratio agrees with the case of interactions between
an old SNR and its ambient GMC \citep[e.g.,][]{2006MNRAS.371.1975Y}. 

In the scenario, the GeV emission from the system may be dim with respect
to the TeV emission as shown in Table 2 of \cite{2006MNRAS.371.1975Y}.
If the contribution of CR protons for the GeV band emission 
can be neglected, the total energy of the accelerated protons for 
the 1--10~TeV emission is estimated to be
$W_p^{\rm total}({\rm 10-100~TeV}) \approx L_{\gamma}({\rm 1-10~TeV}) \tau_{\rm pp}$,
where, $L_{\gamma}({\rm 1-10~TeV})$=4$\pi d^2 F({\rm 1-10~TeV}) \approx 
1.6\times10^{34} (d/4.1~{\rm kpc})^2$~erg~s$^{-1}$ is the luminosity of 
the source in the 1--10~TeV band. Adopting the characteristic cooling time 
of inelastic pp-interactions of 
$\tau_{\rm pp}$=4.4$\times10^{15} (n_{\rm gas}/1~{\rm cm}^{-3})^{-1}$~s
\citep{2004vhec.book.....A} and a mean medium density of 60~cm$^{-3}$ 
for the molecular environment,
$W_p^{\rm total}({\rm 10-100~TeV}) \approx 1.2\times10^{48} (d/4.1~{\rm kpc})^2
(n_{\rm gas}/60~{\rm cm}^{-3})^{-1}$~erg.  
Therefore, an SNR's mechanical explosion energy of 10$^{51}$~erg may
be sufficient enough to explain the observed TeV flux by pp-interactions
in a distance of 4.1~kpc and an average density of $\sim 10^{2}$~cm$^{-3}$.

Finally,
adopting $R_{\rm TeV(1-10~TeV)/radio(10^{7}-10^{11}~Hz)}$=350--50
\citep{2006MNRAS.371.1975Y}
and a non-thermal radio spectral index of $\alpha$=$-$0.5 for
an SNR (e.g., a power-law spectrum of $S(\nu)\propto\nu^{\alpha}$), 
an integral flux from $10^{7}$ to $10^{11}$~Hz of 
$F_{\rm radio}$=2.3$\times10^{-14}$--1.6$\times10^{-13}$~erg~cm$^{-2}$~s$^{-1}$
corresponds to a flux density of 0.1--0.7~Jy at 1.4~GHz.
Such a low radio flux density indicates that the remnant's feature is 
difficult to be discerned from the diffuse background
emission of the Galactic plane
based on the current radio survey \citep[e.g., see Figure 10 in][]{2017arXiv170510927A}.

\section{SUMMARY AND CONCLUSIONS}
We have studied the CO and \mbox{H\,\textsc{i}} data toward the shell-like
TeV source of HESS J1912$+$101. 
The main results and conclusions are summarized as follows.

1. Both of the shocked molecular gas and the high-velocity 
\mbox{H\,\textsc{i}} atomic gas are found to be toward the 
shell-like TeV source of HESS J1912$+$101. The velocity of the 
\twCO\ broadening is obviously up to $V_{\rm LSR}\sim +80\km\ps$
with respect to the unperturbed dense molecular gas at 
$V_{\rm LSR}\sim +60\km\ps$ traced by \thCO\ and C$^{18}$O~($J$=1--0) 
emission.
 
2. The disturbed molecular gas shows compelling signs of the
redshifted-broadening ($\gsim20\km\ps$) in \twCO~($J$=1--0) and 
\twCO~($J$=3--2) lines, which is consistent with the discovery of 
the high-velocity atomic shells up to $V_{\rm LSR}\gsim +100\km\ps$
at the same region. The high-velocity \mbox{H\,\textsc{i}} shells 
are indeed found to be coincident with the molecular partial shells,
suggesting the physical association between them.

3. The shocked molecular gas, the quiescent molecular gas, and the 
high-velocity atomic gas seem to be arranged orderly from 
southeast to northwest, indicating an energetic source with 
large momentum and mechanical energy injection in the southeastern region 
of the disturbed gas concentration.
After excluding other energetic sources such as \HII\ regions and massive
OB stars, we suggest that an old SNR is probably responsible for
the origin of the disturbed gas.

4. We place the old SNR at a near kinematic distance of 4.1 kpc based 
on the association between the remnant and the +60$\km\ps$ MCs. 
The far kinematic distance of the source can be excluded because most 
of the \thCO\ emission from the shocked gas and the molecular shells 
has corresponding \mbox{H\,\textsc{i}} dips at $\sim +60\km\ps$.
The age of the remnant is estimated to be $\sim$(0.7--2.0)$\times10^{5}$~years
according to the evolution of the SNR in a pre-existent
wind bubble from the massive progenitor star.

5. PSR J1913$+$1011, which seems to be just 
located near the centroid of HESS J1912$+$101, probably
results from the core collapse of the massive
progenitor of the old SNR because of their comparable ages
($t_{\rm PSR}\sim$1.7$\times10^{5}$~years vs. 
$t_{\rm SNR}\sim$(0.7--2.0)$\times10^{5}$~years)
and distances ($d_{\rm PSR}\sim$4.5~kpc vs. $d_{\rm SNR}\sim$4.1~kpc). 
The massive progenitor of the remnant blew or carved a molecular bubble 
via its energetic stellar winds, which has a comparable size with
the TeV shell or the +60$\km\ps$ GMC.
Accordingly, the spectral type of its main-sequence star is probably 
earlier than O9V assuming that a massive single star evolved in a 
constant interclump pressure of $p/k\sim 10^{5}$~cm$^{-3}$~K.

6. We suggest that the old SNR may be a promising candidate for the 
shell-like TeV source HESS J1912$+$101, although no corresponding
radio and X-ray counterpart has been detected. The SNR's old age of 
$\sim$(0.7--2.0)$\times10^{5}$~years could be responsible for the absence 
of its radio and X-ray emission.
In the scenario, the shell-like high-energy emission comes from 
the decay of neutral pions produced by interactions of accelerated 
hadrons from the SNR's shock and the surrounding dense gas. 

7. Further high-resolution and sensitivity radio observations
toward the TeV source should be carried out to investigate the
possible non-thermal radio emission of the old SNR.

\acknowledgments
The authors acknowledge the staff members of the Qinghai Radio
Observing Station at Delingha for their support of the observations.
We thank the anonymous referee for valuable comments
and suggestions that helped to improve this paper.
This work is supported in part by National Key R\&D Program of China
(2017YFA0402700).
Y.S. acknowledges support from NSFC grants 11233001 and 11233007. 
X.Z. acknowledges support from NSFC grant 11403104 and Jiangsu Provincial Natural
Science Foundation grant BK20141044. Y.C. acknowledges support from
973 Program grant 2015CB857100 and NSFC grant 11633007.
Y.G. acknowledges support from NSFC grant 11127903.

\facility{PMO 13.7m}
\software{GILDAS/CLASS \citep{2005sf2a.conf..721P}} %\citep{2013ascl.soft05010G}}

\bibliographystyle{aasjournal}
%\bibliography{references}

\clearpage

\begin{deluxetable}{cccccc}
\tabletypesize{\scriptsize}
\tablecaption{Details of the Survey Data Used in the Analysis}
\tablehead{
\colhead{\begin{tabular}{c}
Name \\
   \\
\end{tabular}} &
\colhead{\begin{tabular}{c}
Survey Range         \\
  \\
\end{tabular}} &
\colhead{\begin{tabular}{c}
Line/Continuum\\
  \\
\end{tabular}} &
\colhead{\begin{tabular}{c}
Resolution      \\
 \\
\end{tabular}} &
\colhead{\begin{tabular}{c}
rms sensitivity  \\
 \\
\end{tabular}} &
\colhead{\begin{tabular}{c}
Reference      \\
 \\
\end{tabular}}
}
\startdata
    	          &   								   & 
\twCO~($J$=1--0)  &    50$''$, 0.2~km~s$^{-1}$  & $\sim$0.5~K &       \\
\cline{3-5}
MWISP             &  -10$^{\circ}<l<250^{\circ}$, $\left | b \right | <5^{\circ}$  & 
\thCO~($J$=1--0) & 50$''$, 0.2~km~s$^{-1}$     & $\sim$0.3~K &   (1)  \\
\cline{3-5}
                  &  								   & 
C$^{18}$O~($J$=1--0) & 50$''$, 0.2~km~s$^{-1}$  & $\sim$0.3~K &         \\
\hline
          	  & 10\fdg25$^{\circ}<l<$17\fdg5, $\left | b \right | <$0\fdg5     & 
		  & 			        & 	      & 	 \\
COHRS             & 17\fdg25$^{\circ}<l<$50\fdg25, $\left | b \right | <$0\fdg25   &
\twCO~($J$=3--2)  & 17$''$, 1~km~s$^{-1}$       & $\sim$1~K   &   (2)    \\
         	  & 50\fdg25$<l<$55\fdg25, $\left | b \right | <$0\fdg5    	   &
                  &                        	&     	      & 	  \\
\hline
	          & 18$^{\circ}<l<46^{\circ}$, $\left | b \right | <$1\fdg3        &  
\mbox{H\,\textsc{i}} line & 1$'$, 1.56~km~s$^{-1}$  & $\sim$2~K    &      \\
\cline{3-5}
VGPS              & 46$^{\circ}<l<59^{\circ}$, $\left | b \right | <$1\fdg9        &  
21~cm continuum      	  & 1$'$                    & $\sim$0.3~K  &  (3)  \\
         	  & 59$^{\circ}<l<67^{\circ}$, $\left | b \right | <$2\fdg3        &  
		  &   				    &  		   &   \\
\hline
\enddata
\tablecomments{
(1)~The 34\fdg75$<l<$45\fdg25 and -5\fdg25$<b<$5\fdg25 region has been 
completely mapped \citep{2016ApJ...828...59S}. Also see the information in 
http://english.dlh.pmo.cas.cn/ic/;
(2)~\cite{2013ApJS..209....8D}; 
(3)~\cite{2006AJ....132.1158S}.
%\citep{2013ARAA..51..207B}
}
\end{deluxetable}

\begin{deluxetable*}{cccccccccc}[b!]
\tablecaption{Properties of the High-velocity \mbox{H\,\textsc{i}} Shells}
\tablecolumns{6}
\tablenum{2}
\tablewidth{0pt}
\tablehead{
\colhead{Name} &
\colhead{Size} &
\colhead{$v_{\rm LSR}$\tablenotemark{a}} & 
\colhead{$v_{\rm exp}$\tablenotemark{b}} & 
\colhead{$T_{\rm mean}$\tablenotemark{c}} &
\colhead{Column Density\tablenotemark{d}} &
\colhead{$n_{\rm shell}$\tablenotemark{e,f}} &
\colhead{\mbox{H\,\textsc{i}} Mass\tablenotemark{f}} &
\colhead{$E_{\rm kin}$\tablenotemark{f}} \\
\colhead{}              & \colhead{(arcmin$\times$arcmin)} & \colhead{($\km\ps$)} & 
\colhead{($\km\ps$)}    & \colhead{(K)}      &  \colhead{($10^{20}$~cm$^{-2}$)}     &
\colhead{(cm$^{-3}$)}   & \colhead{($M_{\odot}$)} & \colhead{($\times10^{48}$~ergs)} 
}
\startdata
Shell-a & 3.3$\times$20.4 & 60 & 40 & 0.9 & 0.7 & 5.8$d_{4.1}^{-1}$ & 
70$d_{4.1}^2$ & 1.1$d_{4.1}^2$ \\
Shell-b & 4.7$\times$24.6 & 61 & 40 & 1.2 & 0.9 & 5.2$d_{4.1}^{-1}$ & 
160$d_{4.1}^2$ & 2.6$d_{4.1}^2$ \\
\enddata
\tablenotetext{a}{The LSR velocity of the quiescent gas from the \thCO\ 
and/or C$^{18}$O emission.}
\tablenotetext{b}{The expansion velocity from the difference between
the mean velocity of the shocked gas and the velocity of the quiescent gas.}
\tablenotetext{c}{The mean temperature of the \mbox{H\,\textsc{i}} shell at
100$\km\ps$ after subtracting the surrounding background of $\sim$0.5~K.}
\tablenotetext{d}{The column density calculated from 
1.822$\times10^{18} \times T_{\rm mean} \times v_{\rm exp}$ cm$^{-2}$.}
\tablenotetext{e}{Assuming a depth of 3.3 and 4.7 arcmin (the mean width of 
the arc) for the \mbox{H\,\textsc{i}}
Shell-a and the \mbox{H\,\textsc{i}} Shell-b, respectively.}
\tablenotetext{f}{Parameter $d_{4.1}$ is the distance to the cloud 
in units of 4.1~kpc (see Section 3).}
\end{deluxetable*}

\end{document}